\documentclass{PoS}

\usepackage{amsmath}
\usepackage{color}
\usepackage{graphicx}
\usepackage{axodraw} 
\usepackage{dcolumn}
\usepackage{bm}
\usepackage{wrapfig}

\newcommand{\iu}{{\rm i}}

\newcolumntype{d}[1]{D{.}{\cdot}{#1}}
\newcolumntype{.}{D{.}{.}{-1}}
\newcolumntype{,}{D{,}{,}{2}}

\title{On calculating disconnected-type hadronic light-by-light scattering diagrams from lattice QCD}

\ShortTitle{On calculating disconnected-type hadronic light-by-light scattering diagrams from lattice QCD}

\author{\speaker{M.~Hayakawa}%
        \\
        Department of Physics, Nagoya University, Japan\\
	Nishina Center, RIKEN, Japan
}

\author{T.~Blum\\
        Physics Department, Connecticut University, USA\\
	RIKEN-BNL Research Center, Brookhaven National Laboratory, USA
}

\author{N.~H.~Christ\\
        Physics Department, Columbia University, USA
}

\author{T.~Izubuchi\\
        Physics Department, Brookhaven National Laboratory, USA \\
        RIKEN-BNL Research Center, Brookhaven National Laboratory, USA
}

\author{L.~C.~Jin\\
        Physics Department, Columbia University, USA
}

\author{C.~Lehner\\
        Physics Department, Brookhaven National Laboratory, USA
}

\abstract{
 For reliable comparison of 
the standard model prediction to the muon $g-2$ with
its experimental value, 
the hadronic light-by-light scattering (HLbL) contribution 
must be calculated by lattice QCD simulation. 
 HLbL contribution has many types of disconnected-type diagrams.
 Here, we start with recalling the point that must be taken care of
in every method to calculate them by lattice QCD, 
and present  one concrete method called nonperturbative QED method.
}

\FullConference{The 33rd International Symposium on Lattice Field Theory\\
                 14 -18 July  2015\\
                 Kobe International Conference Center, Kobe, Japan}

\begin{document}

\section{Introduction}

 The hadronic light-by-light scattering (HLbL) contribution,
$a_\mu({\rm HLbL})$, will leave a controversial uncertainty
in the standard model prediction $a_\mu({\rm th})$ to the muon $g-2$, 
unless it can be calculated by means of lattice QCD simulation.

\begin{wraptable}[8]{r}{0.34\linewidth}
 \vspace*{-\intextsep} 
 \caption{
 Comparison of the discrepancy between theory and experiment
with HLbL contribution.
  All are given in units of $10^{-11}$.
}
 \label{tab:muon_g-2}
 \begin{tabular}{rr}
  \hline
  \hline
   $a_\mu({\rm exp}) - a_\mu({\rm th})$ 
   & 
   $249\ (87)$ \\
  \hline
   $a_\mu({\rm HLbL})$ 
   & 
   $116\ (40)$ \\
  \hline
  $\delta a_\mu({\rm next\ exp})$
  & $O(1)$ \\
  \hline
 \end{tabular}
\end{wraptable}

 The reason is as follows.
 While the hadronic vacuum polarization (HVP) contribution
to the muon $g-2$ can resort to the experiments to evaluate 
the relevant QCD dynamics, 
$a_\mu({\rm HLbL})$ requires purely theoretical consideration.
 Thus far, it has been only estimated according
to the models with several hadrons such as pions as dynamical variables.
 The value of $a_\mu({\rm HLbL})$ in Tab.~\ref{tab:muon_g-2} 
was obtained as such \cite{Prades:2009tw}.
 Including it as the part of 
$a_\mu({\rm th})$, 
we observe the discrepancy between 
the experiment and $a_\mu({\rm th})$, which is comparable in size
with $a_\mu({\rm HLbL})$. 
 Actually, no proof supporting the validity of the low energy approximation
to $a_\mu({\rm HLbL})$ exists.
 There is thus a potential possibility of
significance of QCD dynamics that cannot be captured by hadron models.
 Therefore, the first-principle calculation with quarks and gluon 
as dynamical variables, such as lattice QCD simulation, 
is crucial to provide $a_\mu({\rm HLbL})$ with manageable theoretical
uncertainty.		 

\begin{figure}[h]
\begin{center}
\begin{picture}(80,65)(0,0)
\SetWidth{1.2} 
\SetColor{Black} 
\ArrowArcn(35,35)(15,90,270) 
\ArrowArcn(35,35)(15,270,90) 
\SetWidth{1.2} 
\SetColor{Blue} 
\ArrowLine(5,5)(25,5) 
\Line(25,5)(50,5) 
\ArrowLine(50,5)(65,5)
\Text(72,8)[t]{\scriptsize $\mu$}
\SetWidth{1.2} 
\SetColor{Black} 
\Photon(35,50)(35,60){1.5}{2.5} 
\SetColor{Green} 
\Vertex(35,50){1.8} 
\SetWidth{1.2} 
\SetColor{Black} 
\Photon(24,24)(24,5){1.5}{4}
\SetColor{Green} 
\Vertex(24,24){1.8} 
\Vertex(24,5){1.8}
%
\SetColor{Black} 
\Photon(35,20)(35,5){1.5}{4}
\SetColor{Green} 
\Vertex(35,20){1.8} 
\Vertex(35,5){1.8}
%
\SetColor{Black} 
\Photon(46,24)(46,5){1.5}{4} 
\SetColor{Green} 
\Vertex(46,24){1.8} 
\Vertex(46,5){1.8}
\SetWidth{1.0}
\SetColor{Black}
\Line(17,50)(12,35)
\Line(12,35)(17,20)
\Line(52,50)(57,35)
\Line(57,35)(52,20)
\Text(63,22)[t]{\scriptsize \rm QCD}
\end{picture}
\begin{picture}(220,30)(0,0)
\SetColor{Black}
\Text(75,20)[b]{+ 5 permutations of QED vertices (}
\SetColor{Green}
\Vertex(154,25){1.8}
\SetColor{Black}
\Text(198,20)[b]{) on the muon side}
\end{picture}
\caption{
 Connected-type HLbL diagrams.
 Each quark line under the QCD average represents
the inverse of the quark Dirac operator ${\sf D}[U]$
for a given QCD configuration $U$.
 The diagrams with $O(a)$ local QED vertices are not shown here.
}
\end{center}
\label{fig:connected-type_HLbL}
\vspace*{-\intextsep} 
\end{figure}
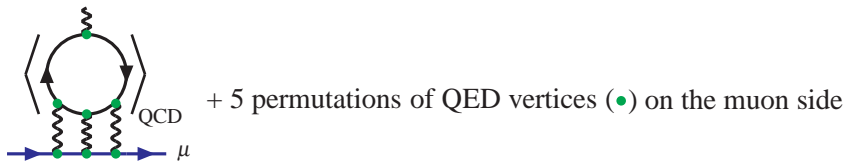

 Recently,
feasibility was demonstrated 
to compute the HLbL contribution 
by the lattice simulation \cite{Blum:2014oka}, 
and more efficient method is investigated in Ref.~\cite{Jin:2015eua}.
 Mainz group has attempted to calculate the HLbL amplitude \cite{Green:2015sra}.
 All of those works, however, focus on 
so-called connected-type diagram shown in Fig.~\ref{fig:connected-type_HLbL}, 
where all of four electromagnetic (EM) vertices lie on a single quark loop.

 Here we turn our attention to 
the disconnected-type HLbL contribution
whose details are presented in Sec.~\ref{sec:classification}.
 We first see the point that must be called into account
in every method to compute it by lattice simulation 
in Sec.~\ref{sec:disconnedctedComponent}.
 In Sec.~\ref{sec:NQEDmethod}, we also present one concrete method
with such a point taken into account
by remedying the one proposed in Ref.~\cite{Blum:2014ita}.

\section{Classification of disconnected-type HLbL diagrams}
\label{sec:classification}

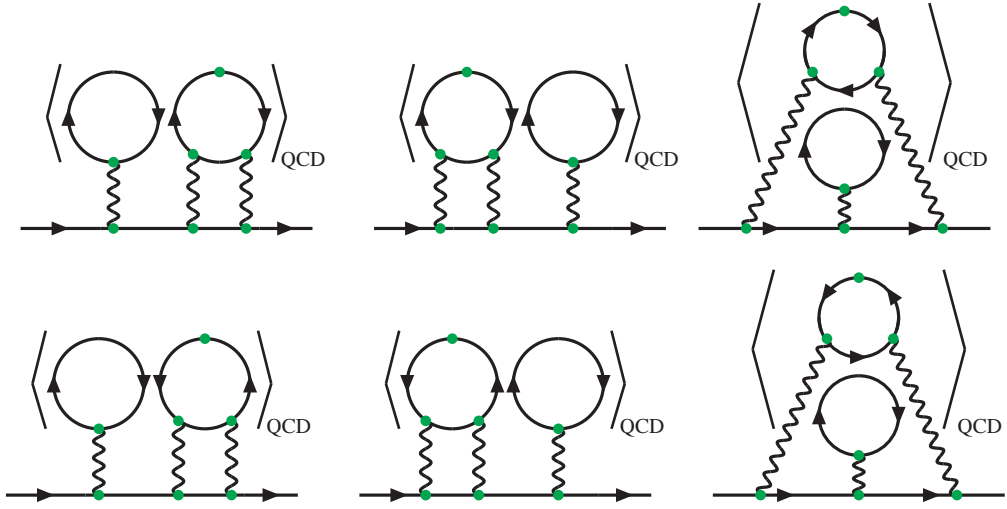
\begin{figure}[h]
\begin{center}
\begin{picture}(120,80)(0,0)
\SetWidth{1.2}
\SetColor{Black}
\ArrowLine(5,5)(35,5)
\Line(35,5)(95,5)
\ArrowLine(95,5)(115,5)
\SetColor{Black}
\ArrowArcn(80,47)(17,90,270)
\ArrowArcn(80,47)(17,270,90)
\SetColor{Green}
\Vertex(80,64){2}
\SetColor{Black}
\Photon(70,33)(70,5){2}{4}
\SetColor{Green}
\Vertex(70,33){2}
\Vertex(70,5){2}
\SetColor{Black}
\Photon(90,33)(90,5){2}{4}
\SetColor{Green}
\Vertex(90,33){2}
\Vertex(90,5){2}
%
\SetColor{Black}
\ArrowArcn(40,47)(17,270,90)
\ArrowArcn(40,47)(17,90,270)
\SetColor{Black}
\Photon(40,30)(40,5){2}{4}
\SetColor{Green}
\Vertex(40,30){2}
\Vertex(40,5){2}
\SetColor{Black}
\SetWidth{1.0}
\Line(20,67)(15,47)
\Line(15,47)(20,30)
\Line(100,67)(105,47)
\Line(105,47)(100,30)
\Text(112,34)[t]{\scriptsize \rm QCD}
\end{picture}
\quad
\begin{picture}(120,80)(0,0)
\SetWidth{1.2}
\SetColor{Black}
\ArrowLine(5,5)(35,5)
\Line(35,5)(95,5)
\ArrowLine(95,5)(115,5)
\SetColor{Black}
\ArrowArcn(40,47)(17,90,270)
\ArrowArcn(40,47)(17,270,90)
\SetColor{Green}
\Vertex(40,64){2}
\SetColor{Black}
\Photon(30,33)(30,5){2}{4}
\SetColor{Green}
\Vertex(30,33){2}
\Vertex(30,5){2}
\SetColor{Black}
\Photon(50,33)(50,5){2}{4}
\SetColor{Green}
\Vertex(50,33){2}
\Vertex(50,5){2}
%
\SetColor{Black}
\ArrowArcn(80,47)(17,270,90)
\ArrowArcn(80,47)(17,90,270)
\SetColor{Black}
\Photon(80,30)(80,5){2}{4}
\SetColor{Green}
\Vertex(80,30){2}
\Vertex(80,5){2}
\SetColor{Black}
\SetWidth{1.0}
\Line(20,67)(15,47)
\Line(15,47)(20,30)
\Line(100,67)(105,47)
\Line(105,47)(100,30)
\Text(112,34)[t]{\scriptsize \rm QCD}
\end{picture}
\begin{picture}(120,100)(0,0)
\SetWidth{1.2}
\SetColor{Black}
\ArrowLine(5,5)(60,5)
\ArrowLine(60,5)(115,5)
\SetColor{Black}
\ArrowArcn(60,72)(15,360,180)
\ArrowArcn(60,72)(15,200,90)
\ArrowArcn(60,72)(15,90,-20)
\SetColor{Green}
\Vertex(60,87){2}
\SetColor{Black}
\Photon(48,64)(23,5){2}{10}
\SetColor{Green}
\Vertex(48,64){2}
\Vertex(23,5){2}
\SetColor{Black}
\Photon(72,64)(97,5){2}{10}
\SetColor{Green}
\Vertex(73,64){2}
\Vertex(97,5){2}
\SetColor{Black}
\ArrowArcn(60,35)(15,90,270)
\ArrowArcn(60,35)(15,270,450)
\SetColor{Black}
\Photon(60,20)(60,5){2}{3}
\SetColor{Green}
\Vertex(60,20){2}
\Vertex(60,5){2}
%
\SetColor{Black}
\SetWidth{1.0}
\Line(28,90)(20,60)
\Line(20,60)(28,30)
\Line(92,90)(100,60)
\Line(100,60)(92,30)
\Text(106,34)[t]{\scriptsize \rm QCD}
\end{picture} 
\\
\begin{picture}(120,80)(0,0)
\SetWidth{1.2}
\SetColor{Black}
\ArrowLine(5,5)(35,5)
\Line(35,5)(95,5)
\ArrowLine(95,5)(115,5)
\SetColor{Black}
\ArrowArc(80,47)(17,90,270)
\ArrowArc(80,47)(17,270,90)
\SetColor{Green}
\Vertex(80,64){2}
\SetColor{Black}
\Photon(70,33)(70,5){2}{4}
\SetColor{Green}
\Vertex(70,33){2}
\Vertex(70,5){2}
\SetColor{Black}
\Photon(90,33)(90,5){2}{4}
\SetColor{Green}
\Vertex(90,33){2}
\Vertex(90,5){2}
%
\SetColor{Black}
\ArrowArcn(40,47)(17,270,90)
\ArrowArcn(40,47)(17,90,270)
\SetColor{Black}
\Photon(40,30)(40,5){2}{4}
\SetColor{Green}
\Vertex(40,30){2}
\Vertex(40,5){2}
\SetColor{Black}
\SetWidth{1.0}
\Line(20,67)(15,47)
\Line(15,47)(20,30)
\Line(100,67)(105,47)
\Line(105,47)(100,30)
\Text(112,34)[t]{\scriptsize \rm QCD}
\end{picture}
\quad
\begin{picture}(120,80)(0,0)
\SetWidth{1.2}
\SetColor{Black}
\ArrowLine(5,5)(35,5)
\Line(35,5)(95,5)
\ArrowLine(95,5)(115,5)
\SetColor{Black}
\ArrowArc(40,47)(17,90,270)
\ArrowArc(40,47)(17,270,90)
\SetColor{Green}
\Vertex(40,64){2}
\SetColor{Black}
\Photon(30,33)(30,5){2}{4}
\SetColor{Green}
\Vertex(30,33){2}
\Vertex(30,5){2}
\SetColor{Black}
\Photon(50,33)(50,5){2}{4}
\SetColor{Green}
\Vertex(50,33){2}
\Vertex(50,5){2}
%
\SetColor{Black}
\ArrowArcn(80,47)(17,270,90)
\ArrowArcn(80,47)(17,90,270)
\SetColor{Black}
\Photon(80,30)(80,5){2}{4}
\SetColor{Green}
\Vertex(80,30){2}
\Vertex(80,5){2}
\SetColor{Black}
\SetWidth{1.0}
\Line(20,67)(15,47)
\Line(15,47)(20,30)
\Line(100,67)(105,47)
\Line(105,47)(100,30)
\Text(112,34)[t]{\scriptsize \rm QCD}
\end{picture}
\quad
\begin{picture}(120,100)(0,0)
\SetWidth{1.2}
\SetColor{Black}
\ArrowLine(5,5)(60,5)
\ArrowLine(60,5)(115,5)
\SetColor{Black}
\ArrowArc(60,72)(15,-20,90)
\ArrowArc(60,72)(15,90,200)
\ArrowArc(60,72)(15,180,360)
\SetColor{Green}
\Vertex(60,87){2}
\SetColor{Black}
\Photon(48,64)(23,5){2}{10}
\SetColor{Green}
\Vertex(48,64){2}
\Vertex(23,5){2}
\SetColor{Black}
\Photon(72,64)(97,5){2}{10}
\SetColor{Green}
\Vertex(73,64){2}
\Vertex(97,5){2}
\SetColor{Black}
\ArrowArcn(60,35)(15,90,270)
\ArrowArcn(60,35)(15,270,450)
\SetColor{Black}
\Photon(60,20)(60,5){2}{3}
\SetColor{Green}
\Vertex(60,20){2}
\Vertex(60,5){2}
%
\SetColor{Black}
\SetWidth{1.0}
\Line(28,90)(20,60)
\Line(20,60)(28,30)
\Line(92,90)(100,60)
\Line(100,60)(92,30)
\Text(106,34)[t]{\scriptsize \rm QCD}
\end{picture} 
\caption{\rm $(3_E,\,1)$-type diagrams.
The diagrams with $O(a)$ local QED vertices are not shown.}
\label{fig:(3E,1):all}
\end{center}
\vspace*{-\intextsep} 
\end{figure}

\begin{figure}[h]
\begin{center}
\begin{picture}(85,100)(0,0)
\SetWidth{1.2}
\SetColor{Black}
\ArrowLine(5,5)(30,5)
\Line(30,5)(50,5)
\ArrowLine(50,5)(75,5)
\SetColor{Black}
\ArrowArcn(40,75)(15,-90,90)
\ArrowArcn(40,75)(15,90,-90)
\SetColor{Green}
\Vertex(40,90){2}
%
\SetColor{Black}
\ArrowArcn(40,40)(15,90,-90)
\ArrowArcn(40,40)(15,-90,90)
\SetColor{Black}
\Photon(29,30)(29,5){2}{4}
\SetColor{Green}
\Vertex(29,30){2}
\Vertex(28, 5){2}
\SetColor{Black}
\Photon(40,25)(40,5){2}{3.5}
\SetColor{Green}
\Vertex(40,25){2}
\Vertex(41, 5){2}
\SetColor{Black}
\Photon(52,30)(52,5){2}{4}
\SetColor{Green}
\Vertex(52,30){2}
\Vertex(51, 5){2}
%
\SetColor{Black}
\SetWidth{1.0}
\Line(22,90)(15,57)
\Line(15,57)(22,24)
\Line(58,90)(65,57)
\Line(65,57)(58,24)
\Text(70,28)[t]{\scriptsize \rm QCD}
\end{picture} 
\quad
\begin{picture}(85,100)(0,0)
\SetWidth{1.2}
\SetColor{Black}
\ArrowLine(5,5)(30,5)
\Line(30,5)(50,5)
\ArrowLine(50,5)(75,5)
\SetColor{Black}
\ArrowArcn(40,75)(15,-90,90)
\ArrowArcn(40,75)(15,90,-90)
\SetColor{Green}
\Vertex(40,90){2}
%
\SetColor{Black}
\ArrowArc(40,40)(15,-90,90)
\ArrowArc(40,40)(15,90,-90)
\SetColor{Black}
\Photon(29,30)(29,5){2}{4}
\SetColor{Green}
\Vertex(29,30){2}
\Vertex(28, 5){2}
\SetColor{Black}
\Photon(40,25)(40,5){2}{3.5}
\SetColor{Green}
\Vertex(40,25){2}
\Vertex(41, 5){2}
\SetColor{Black}
\Photon(52,30)(52,5){2}{4}
\SetColor{Green}
\Vertex(52,30){2}
\Vertex(51, 5){2}
%
\SetColor{Black}
\SetWidth{1.0}
\Line(22,90)(15,57)
\Line(15,57)(22,24)
\Line(58,90)(65,57)
\Line(65,57)(58,24)
\Text(70,28)[t]{\scriptsize \rm QCD}
\end{picture} 
\caption{\rm $\left(1_E,\,3\right)$-type diagrams}
\label{fig:(1E,3):all}
\end{center}
\vspace*{-\intextsep} 
\end{figure}
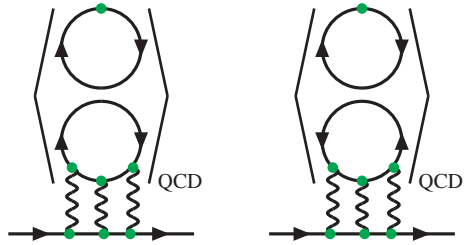

\begin{figure}[h]
\begin{center}
\begin{picture}(120,105)(0,0)
\SetWidth{1.2}
\SetColor{Black}
\ArrowLine(5,5)(115,5)
\SetColor{Black}
\ArrowArcn(60,80)(15,90,270)
\ArrowArcn(60,80)(15,270,90)
\SetColor{Green}
\Vertex(60,95){2}
\SetColor{Black}
\Photon(48,72)(20,5){2}{10}
\SetColor{Green}
\Vertex(48,72){2}
\Vertex(20,5){2}
\SetColor{Black}
\ArrowArcn(60,40)(15,0,180)
\ArrowArcn(60,40)(15,180,0)
\SetColor{Black}
\Photon(47,31)(37,5){2}{4}
\SetColor{Green}
\Vertex(48,30){2}
\Vertex(37,5){2}
\SetColor{Black}
\Photon(71,31)(83,5){2}{4}
\SetColor{Green}
\Vertex(72,30){2}
\Vertex(83,5){2}
%
\SetColor{Black}
\SetWidth{1.0}
\Line(25,98)(15,63)
\Line(15,63)(25,28)
\Line(95,98)(105,63)
\Line(105,63)(95,28)
\Text(108,32)[t]{\scriptsize \rm QCD}
\Text(105,98)[t]{\scriptsize \rm con}
\end{picture} 
\quad
\begin{picture}(120,105)(0,0)
\SetWidth{1.2}
\SetColor{Black}
\ArrowLine(5,5)(115,5)
\SetColor{Black}
\ArrowArcn(60,80)(15,90,270)
\ArrowArcn(60,80)(15,270,90)
\SetColor{Green}
\Vertex(60,95){2}
\SetColor{Black}
\Photon(73,72)(100,5){2}{10}
\SetColor{Green}
\Vertex(73,72){2}
\Vertex(99,5){2}
\SetColor{Black}
\ArrowArcn(60,40)(15,0,180)
\ArrowArcn(60,40)(15,180,0)
\SetColor{Black}
\Photon(47,31)(37,5){2}{4}
\SetColor{Green}
\Vertex(48,30){2}
\Vertex(37,5){2}
\SetColor{Black}
\Photon(71,31)(83,5){2}{4}
\SetColor{Green}
\Vertex(72,30){2}
\Vertex(83,5){2}
%
\SetColor{Black}
\SetWidth{1.0}
\Line(25,98)(15,63)
\Line(15,63)(25,28)
\Line(95,98)(105,63)
\Line(105,63)(95,28)
\Text(108,32)[t]{\scriptsize \rm QCD}
\Text(105,98)[t]{\scriptsize \rm con}
\end{picture} 
\quad
\begin{picture}(120,105)(0,0)
\SetWidth{1.2}
\SetColor{Black}
\ArrowLine(5,5)(60,5)
\ArrowLine(60,5)(115,5)
\SetColor{Black}
\ArrowArcn(60,80)(15,0,180)
\ArrowArcn(60,80)(15,180,360)
\SetColor{Black}
\Photon(48,73)(22,5){2}{10}
\SetColor{Green}
\Vertex(48,72){2}
\Vertex(21,5){2}
\SetColor{Black}
\Photon(74,74)(100,5){2}{10}
\SetColor{Green}
\Vertex(74,74){2}
\Vertex(100,5){2}
\SetColor{Black}
\ArrowArcn(60,40)(15,90,270)
\ArrowArcn(60,40)(15,270,450)
\SetColor{Green}
\Vertex(60,55){2}
\SetColor{Black}
\Photon(60,25)(60,5){2}{3}
\SetColor{Green}
\Vertex(60,25){2}
\Vertex(60,5){2}
%
\SetColor{Black}
\SetWidth{1.0}
\Line(25,98)(15,63)
\Line(15,63)(25,28)
\Line(95,98)(105,63)
\Line(105,63)(95,28)
\Text(108,32)[t]{\scriptsize \rm QCD}
\Text(105,98)[t]{\scriptsize \rm con}
\end{picture} 
\caption{\rm $(2_E,\,2)$-type diagrams}
\label{fig:(2E,2):all}
\end{center}
\vspace*{-\intextsep} 
\end{figure}
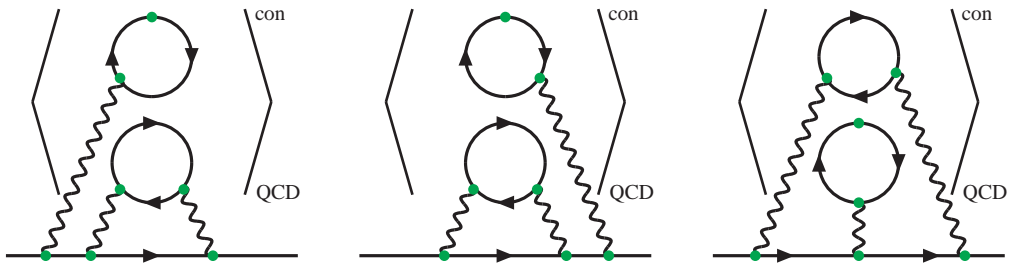

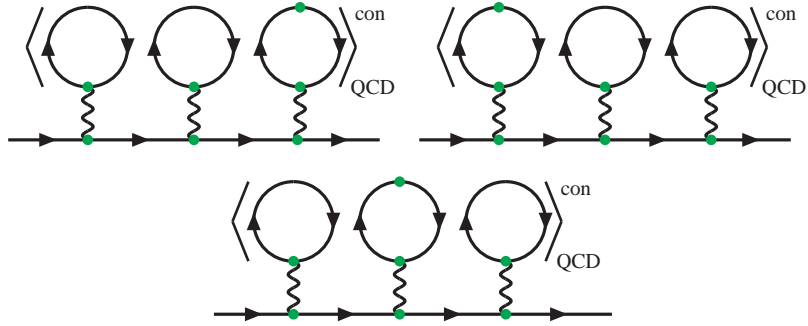
\begin{figure}[h]
\begin{center}
\begin{picture}(150,65)(0,0)
\SetWidth{1.2}
\SetColor{Black}
\ArrowLine(5,5)(35,5)
\ArrowLine(35,5)(75,5)
\ArrowLine(75,5)(115,5)
\ArrowLine(115,5)(145,5)
\SetColor{Black}
\ArrowArcn(115,40)(15,90,270)
\ArrowArcn(115,40)(15,270,90)
\SetColor{Green}
\Vertex(115,55){2}
\SetColor{Black}
\Photon(115,25)(115,5){2}{3}
\SetColor{Green}
\Vertex(115,25){2}
\Vertex(114,5){2}
\SetColor{Black}
\ArrowArcn(75,40)(15,90,-90)
\ArrowArcn(75,40)(15,-90,90)
\SetColor{Black}
\Photon(75,25)(75,5){2}{3}
\SetColor{Green}
\Vertex(75,25){2}
\Vertex(75,5){2}
\SetColor{Black}
\ArrowArcn(35,40)(15,90,-90)
\ArrowArcn(35,40)(15,-90,90)
\SetColor{Black}
\Photon(35,25)(35,5){2}{3}
\SetColor{Green}
\Vertex(35,25){2}
\Vertex(35,5){2}
\SetColor{Black}
\SetWidth{1.0}
\Line(18,55)(12,40)
\Line(12,40)(18,25)
\Line(130,55)(136,40)
\Line(136,40)(130,25)
\Text(143,28)[t]{\scriptsize \rm QCD}
\Text(142,54)[t]{\scriptsize \rm con}
\end{picture}
\ 
\begin{picture}(150,65)(0,0)
\SetWidth{1.2}
\SetColor{Black}
\ArrowLine(5,5)(35,5)
\ArrowLine(35,5)(75,5)
\ArrowLine(75,5)(115,5)
\ArrowLine(115,5)(145,5)
\SetColor{Black}
\ArrowArcn(35,40)(15,90,270)
\ArrowArcn(35,40)(15,270,90)
\SetColor{Green}
\Vertex(35,55){2}
\SetColor{Black}
\Photon(35,25)(35,5){2}{3}
\SetColor{Green}
\Vertex(35,25){2}
\Vertex(35,5){2}
\SetColor{Black}
\ArrowArcn(75,40)(15,90,-90)
\ArrowArcn(75,40)(15,-90,90)
\SetColor{Black}
\Photon(75,25)(75,5){2}{3}
\SetColor{Green}
\Vertex(75,25){2}
\Vertex(75,5){2}
\SetColor{Black}
\ArrowArcn(115,40)(15,90,-90)
\ArrowArcn(115,40)(15,-90,90)
\SetColor{Black}
\Photon(115,25)(115,5){2}{3}
\SetColor{Green}
\Vertex(115,25){2}
\Vertex(115,5){2}
\SetColor{Black}
\SetWidth{1.0}
\Line(18,55)(12,40)
\Line(12,40)(18,25)
\Line(130,55)(136,40)
\Line(136,40)(130,25)
\Text(143,28)[t]{\scriptsize \rm QCD}
\Text(142,54)[t]{\scriptsize \rm con}
\end{picture}
\\ 
\begin{picture}(150,65)(0,0)
\SetWidth{1.2}
\SetColor{Black}
\ArrowLine(5,5)(35,5)
\ArrowLine(35,5)(75,5)
\ArrowLine(75,5)(115,5)
\ArrowLine(115,5)(155,5)
\SetColor{Black}
\ArrowArcn(75,40)(15,90,270)
\ArrowArcn(75,40)(15,270,90)
\SetColor{Green}
\Vertex(75,55){2}
\SetColor{Black}
\Photon(75,25)(75,5){2}{3}
\SetColor{Green}
\Vertex(75,25){2}
\Vertex(75,5){2}
\SetColor{Black}
\ArrowArcn(35,40)(15,90,-90)
\ArrowArcn(35,40)(15,-90,90)
\SetColor{Black}
\Photon(35,25)(35,5){2}{3}
\SetColor{Green}
\Vertex(35,25){2}
\Vertex(35,5){2}
\SetColor{Black}
\ArrowArcn(115,40)(15,90,-90)
\ArrowArcn(115,40)(15,-90,90)
\SetColor{Black}
\Photon(115,25)(115,5){2}{3}
\SetColor{Green}
\Vertex(115,25){2}
\Vertex(115,5){2}
\SetColor{Black}
\SetWidth{1.0}
\Line(18,55)(12,40)
\Line(12,40)(18,25)
\Line(130,55)(136,40)
\Line(136,40)(130,25)
\Text(143,28)[t]{\scriptsize \rm QCD}
\Text(142,54)[t]{\scriptsize \rm con}
\end{picture}
\caption{\rm $(2_E,\,1,\,1)$-type diagrams}
\label{fig:(2E,1,1):all}
\end{center}
\vspace*{-\intextsep} 
\end{figure}

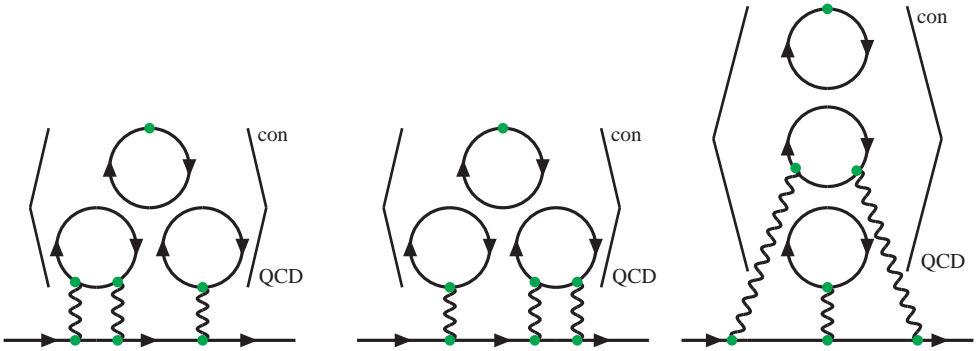
\begin{figure}[h]
\begin{center}
\begin{picture}(120,100)(0,0)
\SetWidth{1.2}
\SetColor{Black}
\ArrowLine(5,5)(40,5)
\ArrowLine(40,5)(80,5)
\ArrowLine(80,5)(115,5)
\SetColor{Black}
\ArrowArcn(60,70)(15,90,270)
\ArrowArcn(60,70)(15,270,90)
\SetColor{Green}
\Vertex(60,85){2}
\SetColor{Black}
\ArrowArcn(40,40)(15,270,90)
\ArrowArcn(40,40)(15,90,-90)
\SetColor{Black}
\Photon(32,27)(32,5){2}{3.5}
\SetColor{Green}
\Vertex(32,27){2}
\Vertex(32,5){2}
\SetColor{Black}
\Photon(48,27)(48,5){2}{3.5}
\SetColor{Green}
\Vertex(48,27){2}
\Vertex(48,5){2}
\SetColor{Black}
\ArrowArcn(80,40)(15,450,270)
\ArrowArcn(80,40)(15,270,90)
\SetColor{Black}
\Photon(80,25)(80,5){2}{3.5}
\SetColor{Green}
\Vertex(80,25){2}
\Vertex(80,5){2}
\SetColor{Black}
\SetWidth{1.0}
\Line(22,85)(15,55)
\Line(15,55)(22,25)
\Line(97,85)(104,55)
\Line(104,55)(97,25)
\Text(110,32)[t]{\scriptsize \rm QCD}
\Text(107,84)[t]{\scriptsize \rm con}
\end{picture}
\quad
\begin{picture}(120,100)(0,0)
\SetWidth{1.2}
\SetColor{Black}
\ArrowLine(5,5)(40,5)
\ArrowLine(40,5)(80,5)
\ArrowLine(80,5)(115,5)
\SetColor{Black}
\ArrowArcn(60,70)(15,90,270)
\ArrowArcn(60,70)(15,270,90)
\SetColor{Green}
\Vertex(60,85){2}
\SetColor{Black}
\ArrowArcn(80,40)(15,270,90)
\ArrowArcn(80,40)(15,90,-90)
\SetColor{Black}
\Photon(72,27)(72,5){2}{3.5}
\SetColor{Green}
\Vertex(72,27){2}
\Vertex(72,5){2}
\SetColor{Black}
\Photon(88,27)(88,5){2}{3.5}
\SetColor{Green}
\Vertex(88,27){2}
\Vertex(88,5){2}
\SetColor{Black}
\ArrowArcn(40,40)(15,450,270)
\ArrowArcn(40,40)(15,270,90)
\SetColor{Black}
\Photon(40,25)(40,5){2}{3.5}
\SetColor{Green}
\Vertex(40,25){2}
\Vertex(40,5){2}
\SetColor{Black}
\SetWidth{1.0}
\Line(22,85)(15,55)
\Line(15,55)(22,25)
\Line(97,85)(104,55)
\Line(104,55)(97,25)
\Text(110,32)[t]{\scriptsize \rm QCD}
\Text(107,84)[t]{\scriptsize \rm con}
\end{picture}
\begin{picture}(120,140)(0,0)
\SetWidth{1.2}
\SetColor{Black}
\ArrowLine(5,5)(30,5)
\Line(30,5)(90,5)
\ArrowLine(90,5)(115,5)
\SetColor{Black}
\ArrowArcn(60,115)(15,90,270)
\ArrowArcn(60,115)(15,270,90)
\SetColor{Green}
\Vertex(60,130){2}
\SetColor{Black}
\ArrowArcn(60,78)(15,90,270)
\ArrowArcn(60,78)(15,270,90)
\SetColor{Black}
\Photon(48,70)(25,5){2}{10}
\SetColor{Green}
\Vertex(48,70){2}
\Vertex(24,5){2}
\SetColor{Black}
\Photon(71,69)(95,5){2}{10}
\SetColor{Green}
\Vertex(71,69){2}
\Vertex(94,5){2}
\SetColor{Black}
\ArrowArcn(60,40)(15,90,270)
\ArrowArcn(60,40)(15,270,90)
\SetColor{Black}
\Photon(60,25)(60,5){2}{3.5}
\SetColor{Green}
\Vertex(60,25){2}
\Vertex(60,5){2}
\SetColor{Black}
\SetWidth{1.0}
\Line(30,131)(17,81)
\Line(17,81)(30,31)
\Line(90,131)(103,81)
\Line(103,81)(90,31)
\Text(104,38)[t]{\scriptsize \rm QCD}
\Text(100,129)[t]{\scriptsize \rm con}
\end{picture}
\caption{\rm $\left(1_E,\,1,\,2\right)$-type diagrams}
\label{fig:(1E,1,2):all}
\end{center}
\vspace*{-\intextsep} 
\end{figure}

\begin{figure}[h]
\begin{center}
\begin{picture}(160,100)(0,0)
\SetWidth{1.2}
\SetColor{Black}
\ArrowLine(5,5)(50,5)
\ArrowLine(50,5)(75,5)
\ArrowLine(75,5)(110,5)
\ArrowLine(110,5)(155,5)
\SetColor{Black}
\ArrowArcn(80,75)(15,90,270)
\ArrowArcn(80,75)(15,270,90)
\SetColor{Green}
\Vertex(80,90){2}
\SetColor{Black}
\ArrowArcn(40,40)(15,90,270)
\ArrowArcn(40,40)(15,270,90)
\SetColor{Black}
\Photon(40,25)(40,5){2}{3.5}
\SetColor{Green}
\Vertex(40,25){2}
\Vertex(40,5){2}
\SetColor{Black}
\ArrowArcn(80,40)(15,90,-90)
\ArrowArcn(80,40)(15,-90,90)
\SetColor{Black}
\Photon(80,25)(80,5){2}{3.5}
\SetColor{Green}
\Vertex(80,25){2}
\Vertex(80,5){2}
\SetColor{Black}
\ArrowArcn(120,40)(15,90,-90)
\ArrowArcn(120,40)(15,-90,90)
\SetColor{Black}
\Photon(120,25)(120,5){2}{3.5}
\SetColor{Green}
\Vertex(120,25){2}
\Vertex(120,5){2}
\SetColor{Black}
\SetWidth{1.0}
\Line(22,95)(12,60)
\Line(12,60)(22,25)
\Line(137,95)(147,60)
\Line(147,60)(137,25)
\Text(150,32)[t]{\scriptsize \rm QCD}
\Text(148,95)[t]{\scriptsize \rm con}
\end{picture}
\caption{\rm $\left(1_E,\,1,\,1,\,1\right)$-type diagrams}
\label{fig:(1E,1,1,1):all}
\end{center}
\vspace*{-\intextsep} 
\end{figure}
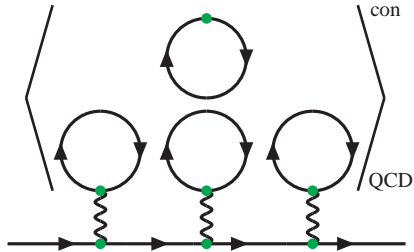

 The lattice QCD simulation tempts us into
classifying the diagrams according to the way how 
four EM vertices are distributed over quark loops.
 It turns out that there are seven types of disconnected-type diagrams in total.
 The diagrams of type $(3_E,\,1)$ 
in Fig.~\ref{fig:(3E,1):all} are those with
three EM vertices on a quark loop, 
one of which couples to the external photon,
and one internal EM vertex on the other loop.
 The diagrams of type $(1_E,\,3)$ in Fig.~\ref{fig:(1E,3):all}
differ from those of type $(3_E,\,1)$ 
because there is no internal EM vertex on the quark loop 
with the external EM vertex.
 The diagrams of type $(2_E,\,2)$ in Fig.~\ref{fig:(2E,2):all}
\footnote{
 The meaning of the superscript ``con'' attached to the average symbol
will be clarified in Sec.~\ref{sec:disconnedctedComponent}.
}
are also the disconnected-type diagrams 
having just two quark loops with two internal EM vertices on each.
 The diagrams in Figs.~\ref{fig:(2E,1,1):all}
and \ref{fig:(1E,1,2):all} are those having three quark loops
with at least one EM vertex on each.
 The diagram in Figs.~\ref{fig:(1E,1,1,1):all}
has four quark loops
with just one EM vertex on each.

\section{Disconnected component in
the correlation function of four EM currents}
\label{sec:disconnedctedComponent}
			 
 Lattice QCD simulation may enable to compute the vacuum expectation value
(VEV) of four hadronic EM currents $j_\mu(x)$
\begin{align}		 
 &			 
 \left<			 
  j_{\mu_{(1)}}(x_{(1)})\,j_{\mu_{(2)}}(x_{(2)})\,j_{\mu_{(3)}}(x_{(3)})\,
  j_{\mu_{(4)}}(x_{(4)}) 
 \right>_{\rm QCD}	 
  \nonumber\\\		 
 &			 
 \quad			 
 =			 
 \frac{1}{Z_{\rm QCD}}	 
 \int dU \int dq d\overline{q}\,
  e^{- S_{\rm QCD}\left[U,\,q,\,\overline{q}\right]}
  j_{\mu_{(1)}}(x_{(1)})\,j_{\mu_{(2)}}(x_{(2)})\,j_{\mu_{(3)}}(x_{(3)})\,
  j_{\mu_{(4)}}(x_{(4)})\,.\label{eq:avr:fourEM}
\end{align}		 
 This VEV, however, contains not only the contribution from
the field-theoretically connected diagrams,
but also that from the field-theoretically disconnected diagrams.
 Here, a field-theoretically disconnected diagram
is a Feynman diagram consisting of
more than one nontrivial connected subgraph as a graph.		 
 We shall refer to the contribution of field-theoretically
disconnected (connected) diagrams as the 
{\it disconnected} ({\it connected}) {\it component}
of that contribution.	 
 If the connected component of the QCD average of $\mathcal{A}$
is denoted by $\left<\mathcal{A}\right>_{\rm QCD}^{\rm con}$, 
the quantity relevant to the HLbL
is 
$			 
\left<			 
  j_{\mu_{(1)}}(x_{(1)})\,j_{\mu_{(2)}}(x_{(2)})\,j_{\mu_{(3)}}(x_{(3)})\,
  j_{\mu_{(4)}}(x_{(4)}) 
\right>_{\rm QCD}^{\rm con}
$, 			 
which differs 		 
from the one in Eq.~(\ref{eq:avr:fourEM}) 
by the sum of three terms each of which is a product 
of two currents, $\left< j_{\lambda}(x)\,j_{\rho}(y) \right>_{\rm QCD}$. 

 Basically,
lattice QCD simulation does not allow us to evaluate
$\left<	\mathcal{A} \right>_{\rm QCD}^{\rm con}$ directly 
because connectivity is the attribute of each Feynman diagram
with quarks and gluons.
 The best we can do is to compute 
$\left<	\mathcal{A} \right>_{\rm QCD}$
and its disconnected component 
$\left<	\mathcal{A} \right>_{\rm QCD}^{\rm dis}$
to get
$\left<	\mathcal{A} \right>_{\rm QCD}^{\rm con}
= \left< \mathcal{A} \right>_{\rm QCD} -
\left<\mathcal{A} \right>_{\rm QCD}^{\rm dis}$ indirectly.

 If the VEV of four EM  currents in Eq.~(\ref{eq:avr:fourEM})
couples to the muon via three virtual photons,
its disconnected component
gives rise to 
the HVP contribution with the $O(\alpha)$ renormalization of EM charge
due to QCD.
 Any calculation of the HVP contribution with the renormalized EM charge 
contains such a contribution implicitly.
 To avoid double counting, we must thus explicitly subtract
such an $O(\alpha^3)$ HVP contribution,
which is henceforth called {\it unwanted contribution}, 
in every method based on lattice QCD.
 Subtraction is required in practice to compute the disconnected-type diagrams; 
$(2_E,\,2)$, $\left(1_E,\,1,\,2\right)$, 
$\left(2_E,\,1,\,1\right)$ and $\left(1_E,\,1,\,1,\,1\right)$
in our classification scheme.

\section{Nonperturbative QED method for full HLbL contribution}
\label{sec:NQEDmethod}

\begin{wraptable}[12]{r}{0.32\linewidth}
\vspace*{-\intextsep} 
\caption{
 Emergence of degeneracy. 
\textcolor[named]{Blue}{$C$}, say, denotes
\textcolor[named]{Blue}{$\mathcal{M}_C - \mathcal{S}_C$}.}
\label{tab:multiplicationFactor}
\begin{tabular}{l|c|c}
\hline
\hline
 & 
 \,$\textcolor[named]{Blue}{C}
     + \textcolor[named]{Green}{{C^\prime}}$\,
 & \ $\textcolor[named]{Magenta}{D}$ \,\\
\hline
$4_E$ & $3$ & $0$ \\
$\left(1_E,3\right)$ & $0$ & $3$ \\
$\left(3_E,1\right)$ & $2$ & $1$ \\
$\left(2_E,2\right)$ & $1$ & $2$ \\
$\left(1_E,1,2\right)$ & $0$ & $3$ \\
$\left(2_E,1,1\right)$ & $1$ & $2$ \\
$\left(1_E,1,1,1\right)$ & $0$ & $3$ \\
\hline
\end{tabular}
\end{wraptable}

\begin{figure}[htb]
\begin{center} 
\begin{picture}(121,65)(0,0)
\Text(7,28)[t]{\scriptsize $\textcolor[named]{Blue}{\mathcal{M}_C} = $}
\SetWidth{1.2}
\SetColor{Black}
\ArrowLine(30,10)(55,10)
\ArrowLine(55,10)(80,10)
\SetWidth{1.2}
\SetColor{Black}
\ArrowArc(55,38)(13,-90, 90)
\ArrowArc(55,38)(13, 90,-90)
\SetWidth{1.2}
\SetColor{Black}
\Photon(55,25)(55,10){1.8}{3.5} 
\Photon(55,51)(55,61){1.8}{2.5} 
\SetColor{Green}
\Vertex(55,25){1.8} 
\Vertex(55,10){1.8} 
\Vertex(55,51){1.8} 
\SetColor{Orange}
\SetWidth{0.8}
\Line(28,45)(23,25)
\Line(23,25)(28, 5)
\Vertex(23,25){0.3}
\Line(81,45)(87,25)
\Line(87,25)(81,5)
\Vertex(87,25){0.3}
\Text(99,7)[t]{\tiny \rm \textcolor[named]{OrangeRed}{QCD\,$+$\,QED}}
\end{picture}
\begin{picture}(117,65)(0,0)
\Text(8,28)[t]{\scriptsize $\textcolor[named]{Blue}{\mathcal{S}_C} = $}
\SetWidth{1.2}
\SetColor{Black}
\ArrowLine(30,10)(55,10)
\ArrowLine(55,10)(80,10)
\SetColor{Orange}
\SetWidth{0.8}
\Line(29,15)(25,10)
\Line(25,10)(29,5)
\Vertex(25,10){0.3}
\Line(81,15)(85,10)
\Line(85,10)(81,5)
\Vertex(85,10){0.3}
\Text(100,7)[t]{\tiny \rm \textcolor[named]{OrangeRed}{QCD\,$+$\,QED}}
\SetWidth{1.2}
\SetColor{Black}
\ArrowArc(55,38)(13,-90, 90)
\ArrowArc(55,38)(13, 90,-90)
\SetColor{Orange}
\SetWidth{0.8}
\Line(41,52)(34,38)
\Line(34,38)(41,24)
\Vertex(34,38){0.3}
\Line(69,52)(76,38)
\Line(76,38)(69,24)
\Vertex(76,38){0.3}
\Text(88,27)[t]{\tiny \rm \textcolor[named]{Blue}{QCD\,$+$\,QED}}
\SetWidth{1.2}
\SetColor{Black}
\Photon(55,25)(55,10){1.8}{3.5}
\Photon(55,51)(55,61){1.8}{2.5} 
\SetColor{Green}
\Vertex(55,25){1.8} 
\Vertex(55,10){1.8} 
\Vertex(55,51){1.8} 
\end{picture}
\\
\begin{picture}(121,55)(0,0)
\Text(7,28)[t]
{\scriptsize $\textcolor[named]{Green}{\mathcal{M}_{C^\prime}} = $}
\SetWidth{1.2}
\SetColor{Black}
\ArrowLine(30,10)(55,10)
\ArrowLine(55,10)(80,10)
\SetWidth{1.2}
\SetColor{Black}
\ArrowArc(55,38)(13,-90, 90)
\ArrowArc(55,38)(13, 90,-90)
\SetWidth{1.2}
\SetColor{Black}
\Photon(55,25)(55,10){1.8}{3.5}
\Photon(55,25)(55,35){1.8}{2.5}
\SetColor{Green}
\Vertex(55,25){1.8} 
\Vertex(55,10){1.8} 
\SetColor{Orange}
\SetWidth{0.8}
\Line(28,45)(23,25)
\Line(23,25)(28, 5)
\Vertex(23,25){0.3}
\Line(81,45)(87,25)
\Line(87,25)(81,5)
\Vertex(87,25){0.3}
\Text(99,7)[t]{\tiny \rm \textcolor[named]{OrangeRed}{QCD\,$+$\,QED}}
\end{picture}
\begin{picture}(117,55)(0,0)
\Text(8,28)[t]
{\scriptsize $\textcolor[named]{Green}{\mathcal{S}_{C^\prime}} = $}
\SetWidth{1.2}
\SetColor{Black}
\ArrowLine(30,10)(55,10)
\ArrowLine(55,10)(80,10)
\SetColor{Orange}
\SetWidth{0.8}
\Line(29,15)(25,10)
\Line(25,10)(29,5)
\Vertex(25,10){0.3}
\Line(81,15)(85,10)
\Line(85,10)(81,5)
\Vertex(85,10){0.3}
\Text(100,7)[t]{\tiny \rm \textcolor[named]{OrangeRed}{QCD\,$+$\,QED}}
\SetWidth{1.2}
\SetColor{Black}
\ArrowArc(55,38)(13,-90, 90)
\ArrowArc(55,38)(13, 90,-90)
\SetColor{Orange}
\SetWidth{0.8}
\Line(41,52)(34,38)
\Line(34,38)(41,24)
\Vertex(34,38){0.3}
\Line(69,52)(76,38)
\Line(76,38)(69,24)
\Vertex(76,38){0.3}
\Text(88,27)[t]{\tiny \rm \textcolor[named]{Blue}{QCD\,$+$\,QED}}
\SetWidth{1.2}
\SetColor{Black}
\Photon(55,25)(55,10){1.8}{3.5}
\Photon(55,25)(55,35){1.8}{2.5}
\SetColor{Green}
\Vertex(55,25){1.8} 
\Vertex(55,10){1.8} 
\end{picture}
\caption{
 The terms 
$\textcolor[named]{Blue}{\mathcal{M}_{C}}$,
$\textcolor[named]{Blue}{\mathcal{S}_{C}}$, 
and $\textcolor[named]{Green}{\mathcal{M}_{C^\prime}}$,
$\textcolor[named]{Green}{\mathcal{S}_{C^\prime}}$
with $O(a)$ QED vertices.
}
\label{fig:C+Cprime}
\end{center}
\vspace*{-\intextsep} 
\end{figure}
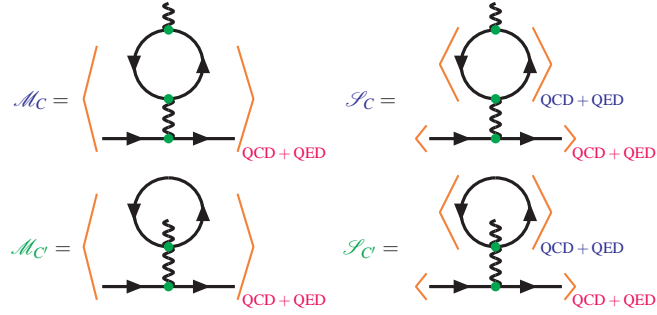

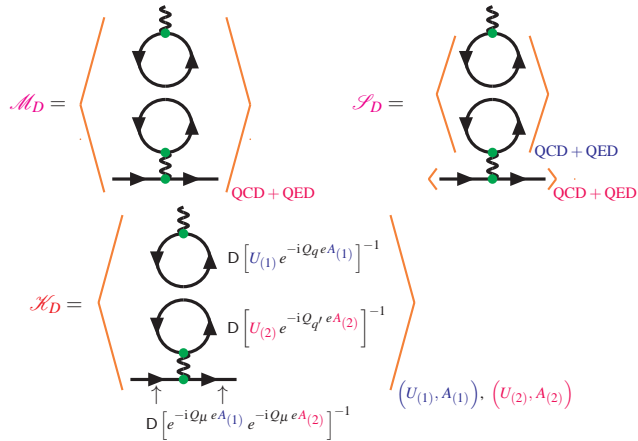
\begin{figure}[htb]
\begin{center}
\begin{picture}(126,80)(0,0)
\Text(7,42)[t]{\scriptsize $\textcolor[named]{Magenta}{\mathcal{M}_D} = $}
\SetWidth{1.2}
\SetColor{Black}
\ArrowLine(35,10)(55,10)
\ArrowLine(55,10)(75,10)
\SetWidth{1.2}
\SetColor{Black}
\ArrowArc(55,30)(10,-90, 90)
\ArrowArc(55,30)(10, 90,-90)
\SetWidth{1.2}
\SetColor{Black}
\ArrowArc(55,55)(10,-90, 90)
\ArrowArc(55,55)(10, 90,-90)
\SetWidth{1.2}
\SetColor{Black}
\Photon(55,65)(55,75){1.8}{2.5}
\SetColor{Green}
\Vertex(55,65){1.8}
\SetWidth{1.2}
\SetColor{Black}
\Photon(55,20)(55,10){1.8}{2.5}
\SetColor{Green}
\Vertex(55,20){1.8}
\Vertex(55,10){1.8}
\SetColor{Orange}
\SetWidth{0.8}
\Line(32,71)(23,38)
\Line(23,38)(32,5)
\Vertex(23,25){0.3}
\Line(78,71)(87,38)
\Line(87,38)(78,5)
\Vertex(87,25){0.3}
\Text(96,7)[t]{\tiny \rm \textcolor[named]{OrangeRed}{QCD\,$+$\,QED}}
\end{picture}
\begin{picture}(125,80)(0,0)
\Text(7,42)[t]{\scriptsize $\textcolor[named]{Magenta}{\mathcal{S}_D} = $}
\SetWidth{1.2}
\SetColor{Black}
\ArrowLine(30,10)(50,10)
\ArrowLine(50,10)(70,10)
\SetColor{Orange}
\SetWidth{0.8}
\Line(29,14)(26,10)
\Line(26,10)(29,6)
\Vertex(26,10){0.3}
\Line(71,14)(74,10)
\Line(74,10)(71,6)
\Vertex(81,10){0.3}
\Text(89,7)[t]{\tiny \rm \textcolor[named]{OrangeRed}{QCD\,$+$\,QED}}
\SetWidth{1.2}
\SetColor{Black}
\ArrowArc(50,30)(10,-90, 90)
\ArrowArc(50,30)(10, 90,-90)
\SetWidth{1.2}
\SetColor{Black}
\ArrowArc(50,55)(10,-90, 90)
\ArrowArc(50,55)(10, 90,-90)
\SetWidth{1.2}
\SetColor{Black}
\Photon(50,65)(50,75){1.8}{2.5}
\SetColor{Green}
\Vertex(50,65){1.8}
\SetWidth{1.2}
\SetColor{Black}
\Photon(50,20)(50,10){1.8}{2.5}
\SetColor{Green}
\Vertex(50,20){1.8}
\Vertex(50,10){1.8}
\SetColor{Orange}
\SetWidth{0.8}
\Line(36,64)(29,42)
\Line(29,42)(36,20)
\Vertex(29,42){0.3}
\Line(64,64)(71,42)
\Line(71,42)(64,20)
\Vertex(71,42){0.3}
\Text(82,23)[t]{\tiny \rm \textcolor[named]{Blue}{QCD\,$+$\,QED}}
\end{picture}
%
\\
\begin{picture}(240,90)(0,0)
\Text(7,57)[t]{\scriptsize $\textcolor[named]{Red}{\mathcal{K}_D} = $}
\SetWidth{1.2}
\SetColor{Black}
\ArrowLine(35,25)(55,25)
\ArrowLine(55,25)(75,25)
\SetColor{Black}
\ArrowArc(55,70)(10,90,270)
\ArrowArc(55,70)(10,270,90)
\SetWidth{1.2}
\SetColor{Black}
\Photon(55,80)(55,90){1.8}{2.5}
\SetColor{Green}
\Vertex(55,80){1.8}
%
\Text(101,78)[t]
{{\tiny ${\sf D}\left[\textcolor[named]{Blue}{U_{(1)}}\,
  e^{-\iu\,Q_q\,e\,\textcolor[named]{Blue}{A_{(1)}}}\right]^{-1}$}}
%
\SetColor{Black}
\ArrowArc(55,45)(10,90,270)
\ArrowArc(55,45)(10,270,90)
\Text(102,53)[t]
{{\tiny ${\sf D}\left[\textcolor[named]{OrangeRed}{U_{(2)}}\,
  e^{-\iu\,Q_{q^\prime}\,e\,\textcolor[named]{OrangeRed}{A_{(2)}}}\right]^{-1}$}}
%
\SetColor{Black}
\Photon(55,35)(55,25){2}{2.5}
\SetColor{Green}
\Vertex(55,35){1.8}
\Vertex(55,25){1.8}
\Text(80,15)[t]
{{\tiny 
${\sf D}\left[e^{-\iu\,Q_{\mu}\,e\,\textcolor[named]{Blue}{A_{(1)}}}\,
e^{-\iu\,Q_{\mu}\,e\,\textcolor[named]{OrangeRed}{A_{(2)}}}\right]^{-1}$}}
%
\Text(45,23)[t]{{\scriptsize $\uparrow$}}
\Text(70,23)[t]{{\scriptsize $\uparrow$}}
%
\SetColor{Orange}
\SetWidth{0.8}
\Line(32,87)(23,54)
\Line(23,54)(32,21)
\Vertex(23,54){0.3}
\Line(133,87)(142,54)
\Line(142,54)(133,21)
\Vertex(142,54){0.3}
\Text(168,25)[t]{
{\tiny \textcolor[named]{Blue}{$\left(U_{(1)},\,A_{(1)}\right)$},\,
\textcolor[named]{OrangeRed}{$\left(U_{(2)},\,A_{(2)}\right)$}}}
\end{picture}
\caption{
 The terms
$\textcolor[named]{Magenta}{\mathcal{M}_{D}}$,
$\textcolor[named]{Magenta}{\mathcal{S}_{D}}$ 
and $\textcolor[named]{Red}{\mathcal{K}_{D}}$.
}
\label{fig:D+KD}
\end{center}
\vspace*{-\intextsep} 
\end{figure}

 The nonperturbative QED method we propose here
to compute full HLbL contribution is given by 
\begin{align}
&
\frac{1}{3}
\left\{
 \left(
  \textcolor[named]{Blue}{\mathcal{M}_C}
  - 
  \textcolor[named]{Blue}{\mathcal{S}_C}
 \right)
 + 
 \left(
  \textcolor[named]{Green}{\mathcal{M}_{C^\prime}}
  - 
  \textcolor[named]{Green}{\mathcal{S}_{C^\prime}}
 \right)
 +
 \left(
  \textcolor[named]{Magenta}{\mathcal{M}_D}
  -
  \textcolor[named]{Magenta}{\mathcal{S}_D}
 \right)
 \textcolor[named]{Red}{- \mathcal{K}_D}
\right\}\,, \label{eq:nonperturbativeQEDMethod}
\end{align}
where the terms $\mathcal{M}_{C}$, $\mathcal{S}_{C}$, 
$\textcolor[named]{Green}{\mathcal{M}_{C^\prime}}$
and $\textcolor[named]{Green}{\mathcal{S}_{C^\prime}}$
are defined in Fig.~\ref{fig:C+Cprime},
and 
$\textcolor[named]{Magenta}{\mathcal{M}_{D}}$,
$\textcolor[named]{Magenta}{\mathcal{S}_{D}}$ 
and $\textcolor[named]{Red}{\mathcal{K}_{D}}$
in Fig.~\ref{fig:D+KD}.
 The individual terms involve the averages with respect to 
(${\rm QCD} + {\rm QED}$) for light quark system.
 Note that each muon line in Figs.~\ref{fig:C+Cprime}
and \ref{fig:D+KD} denotes 
the inverse of muon Dirac operator in the QED configuration 
generated by such (${\rm QCD} + {\rm QED}$) system.
 We multiply $\frac{1}{3}$ to the quantity in the bracket
in Eq.~(\ref{eq:nonperturbativeQEDMethod}) 
because individual $8$ types of HLbL diagrams
emerge with {\it triplicate degeneracy} 
as in Tab.~\ref{tab:multiplicationFactor} 
\cite{Blum:2014ita}.

\begin{figure}[h]
\begin{center}
\begin{picture}(90,83)(0,0)
\SetWidth{1.2}
\SetColor{Black}
\ArrowLine(5,5)(45,5)
\ArrowLine(45,5)(85,5)
\SetColor{Red}
\ArrowArcn(45,67)(12,0,180)
\ArrowArcn(45,67)(12,180,360)
\SetColor{Red}
\Photon(35,60)(16,5){2}{9}
\Vertex(35,60){1.8}
\Vertex(16,5){1.8}
\SetColor{Red}
\Photon(55,60)(74,5){2}{9}
\Vertex(55,60){1.8}
\Vertex(73,5){1.8}
\SetColor{Black}
\ArrowArcn(45,32)(12,90,270)
\ArrowArcn(45,32)(12,270,450)
\SetColor{Black}
\Photon(45,44)(45,34){2}{2.5}
\SetColor{Green}
\Vertex(45,44){1.8}
\SetColor{Black}
\Photon(45,20)(45,5){2}{2.5}
\SetColor{Green}
\Vertex(45,20){1.8}
\Vertex(45,5){1.8}
%
\SetColor{Black}
\SetWidth{1.0}
\Line(15,80)(7,50)
\Line(7,50)(15,20)
\Line(75,80)(83,50)
\Line(83,50)(75,20)
\Text(85,23)[t]{\tiny \rm QCD}
\end{picture} 
\quad
\begin{picture}(90,83)(0,0)
\SetWidth{1.2}
\SetColor{Black}
\ArrowLine(5,5)(45,5)
\ArrowLine(45,5)(85,5)
\SetColor{Black}
\ArrowArcn(45,67)(12,0,180)
\ArrowArcn(45,67)(12,180,360)
\SetColor{Black}
\Photon(35,60)(16,5){2}{9}
\SetColor{Green}
\Vertex(35,60){1.8}
\Vertex(16,5){1.8}
\SetColor{Red}
\Photon(55,60)(74,5){2}{9}
\Vertex(55,60){1.8}
\Vertex(73,5){1.8}
\SetColor{Black}
\ArrowArcn(45,32)(12,90,270)
\ArrowArcn(45,32)(12,270,450)
\SetColor{Black}
\Photon(45,44)(45,34){2}{2.5}
\SetColor{Green}
\Vertex(45,44){1.8}
\SetColor{Red}
\Photon(45,20)(45,5){2}{2.5}
\Vertex(45,20){1.8}
\Vertex(45,5){1.8}
%
\SetColor{Black}
\SetWidth{1.0}
\Line(15,80)(7,50)
\Line(7,50)(15,20)
\Line(75,80)(83,50)
\Line(83,50)(75,20)
\Text(85,23)[t]{\tiny \rm QCD}
\end{picture} 
\quad
\begin{picture}(90,83)(0,0)
\SetWidth{1.2}
\SetColor{Black}
\ArrowLine(5,5)(45,5)
\ArrowLine(45,5)(85,5)
\SetColor{Black}
\ArrowArcn(45,67)(12,0,180)
\ArrowArcn(45,67)(12,180,360)
\SetColor{Red}
\Photon(35,60)(16,5){2}{9}
\Vertex(35,60){1.8}
\Vertex(16,5){1.8}
\SetColor{Black}
\Photon(55,60)(74,5){2}{9}
\SetColor{Green}
\Vertex(55,60){1.8}
\Vertex(73,5){1.8}
\SetColor{Black}
\ArrowArcn(45,32)(12,90,270)
\ArrowArcn(45,32)(12,270,450)
\SetColor{Black}
\Photon(45,44)(45,34){2}{2.5}
\SetColor{Green}
\Vertex(45,44){1.8}
\SetColor{Red}
\Photon(45,20)(45,5){2}{2.5}
\Vertex(45,20){1.8}
\Vertex(45,5){1.8}
%
\SetColor{Black}
\SetWidth{1.0}
\Line(15,80)(7,50)
\Line(7,50)(15,20)
\Line(75,80)(83,50)
\Line(83,50)(75,20)
\Text(85,23)[t]{\tiny \rm QCD}
\end{picture} 
\caption{An identical diagram of $\left(2_E,\,2\right)$-type is generated
in three ways from $\textcolor[named]{Blue}{\mathcal{M}_C}$ (left)
and $\textcolor[named]{Magenta}{\mathcal{M}_D}$ (middle,\,right).
 \textcolor[named]{Red}
{The red propagators and vertices are generated by 
the ensemble average of (${\rm QCD} + {\rm QED}$)
}.
}
\label{fig:(2E,2)}
\end{center}
\vspace*{-\intextsep} 
\end{figure}
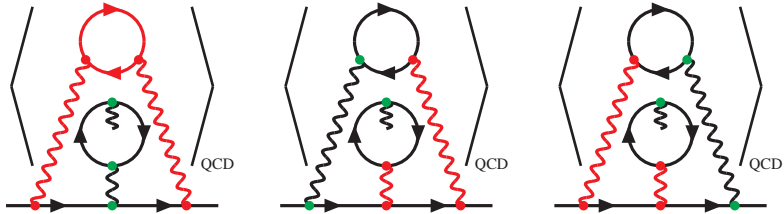

 The term $\left(\textcolor[named]{Red}{-\mathcal{K}_{D}}\right)$
in Eq.~(\ref{eq:nonperturbativeQEDMethod})
is the one added here
to subtract the unwanted $O(\alpha^3)$ HVP contribution
contained in the other terms.
 To construct $\textcolor[named]{Red}{\mathcal{K}_{D}}$, 
we prepare two sets of $\left({\rm QCD},\,{\rm QED}\right)$.
 Practically, they may be two independent important samples of 
a pair of $\left(U,\,A\right)$ 
generated by dynamical (${\rm QCD} + {\rm QED}$) simulation.
 The quark in the upper loop  on the right-hand side of
Fig.~\ref{fig:D+KD} is charged only with respect to
the first $\left({\rm QCD},\,{\rm QED}\right)$ 
and the one in the lower loop only with respect to
the second $\left({\rm QCD},\,{\rm QED}\right)$,
while the muon is charged with respect to both QEDs.

\begin{figure}[htb]
\begin{center}
\begin{picture}(195,30)(0,0)
\SetWidth{1.2}
\SetColor{Black}
\COval(26,15)(12,15)(0){Yellow}{Yellow}
\Text(26,18)[t]{\scriptsize {\rm QCD}}
\SetColor{Black}
\Photon(0,15)(10,15){2}{2}
\SetColor{Red}
\Photon(42,15)(52,15){2}{2}
\SetColor{Green}
\Vertex(10,15){1.5}
\SetColor{Red}
\Vertex(42,15){1.5}
\Text(62,15)[t]{$=$}
\SetWidth{1.2}
\SetColor{Black}
\ArrowArcn(90,15)(8,0,180)
\ArrowArcn(90,15)(8,180,360)
\SetWidth{1.2}
\SetColor{Black}
\Photon(82,15)(72,15){2}{2.5}
\SetColor{Red}
\Photon(98,15)(108,15){2}{2.5}
\SetColor{Green}
\Vertex(82,15){1.5}
\SetColor{Red}
\Vertex(98,15){1.5}
\SetColor{Orange}
\SetWidth{1.0}
\Line(82,25)(78,15)
\Line(78,15)(82, 5)
\Line(98,25)(102,15)
\Line(102,15)(98, 5)
\Text(108,8)[t]{\tiny \rm QCD}
\Text(122,18)[t]{$+$}
\SetWidth{1.2}
\SetColor{Black}
\ArrowArcn(148,15)(8,0,180)
\ArrowArcn(148,15)(8,180,360)
\SetWidth{1.2}
\SetColor{Black}
\Photon(140,15)(130,15){2}{2.5}
\SetColor{Green}
\Vertex(140,15){1.5}
\SetWidth{1.2}
\SetColor{Red}
\ArrowArcn(168,15)(8,0,180)
\ArrowArcn(168,15)(8,180,360)
\SetWidth{1.2}
\SetColor{Red}
\Photon(176,15)(186,15){2}{2.5}
\SetColor{Red}
\Vertex(176,15){1.5}
\SetColor{Orange}
\SetWidth{1.0}
\Line(140,25)(136,15)
\Line(136,15)(140, 5)
\Line(176,25)(180,15)
\Line(180,15)(176, 5)
\Text(186,8)[t]{\tiny \rm QCD}
\end{picture}
\caption{Full HVP function.
The diagrams with $O(a)$ {\rm QED} vertices are not shown.}
\label{fig:fullHVP}
\end{center}
\vspace*{-\intextsep} 
\end{figure}

\begin{figure}[h]
\begin{center}
\begin{picture}(92,90)(0,0)
\SetWidth{1.2}
\SetColor{Black}
\ArrowLine(5,35)(70,35)
\ArrowLine(70,35)(85,35)
\SetColor{Black}
\COval(70,62)(12,12)(0){Yellow}{Yellow}
\Text(70,64)[t]{\tiny \rm QCD}
\SetWidth{1.2}
\SetColor{Black}
\Photon(70,74)(70,84){1.8}{2.5}
\SetColor{Green}
\Vertex(70,75){1.8}
%
\SetColor{Red}
\Photon(70,50)(70,35){2}{3}
\Vertex(70,50){1.8}
\Vertex(70,35){1.8}
%
\SetColor{Black}
\SetColor{Black}
\COval(35,15)(12,12)(0){Yellow}{Yellow}
\Text(35,17)[t]{\tiny \rm QCD}
%
\SetColor{Red}
\Photon(15,35)(25,23){2}{3}
\SetColor{Red}
\Vertex(15,35){1.8}
\Vertex(25,23){1.8}
%
\SetColor{Black}
\Photon(55,35)(45,23){2}{3}
\SetColor{Green}
\Vertex(55,35){1.8}
\Vertex(45,23){1.8}
\end{picture}
\begin{picture}(92,90)(0,0)
\SetWidth{1.2}
\SetColor{Black}
\ArrowLine(5,35)(70,35)
\ArrowLine(70,35)(85,35)
\SetColor{Black}
\COval(70,62)(12,12)(0){Yellow}{Yellow}
\Text(70,64)[t]{\tiny \rm QCD}
\SetWidth{1.2}
\SetColor{Black}
\Photon(70,74)(70,84){1.8}{2.5}
\SetColor{Green}
\Vertex(70,75){1.8}
%
\SetColor{Red}
\Photon(70,50)(70,35){2}{3}
\Vertex(70,50){1.8}
\Vertex(70,35){1.8}
%
\SetColor{Black}
\SetColor{Black}
\COval(35,15)(12,12)(0){Yellow}{Yellow}
\Text(35,17)[t]{\tiny \rm QCD}
%
\SetColor{Black}
\Photon(15,35)(25,23){2}{3}
\SetColor{Green}
\Vertex(15,35){1.8}
\Vertex(25,23){1.8}
%
\SetColor{Red}
\Photon(55,35)(45,23){2}{3}
\Vertex(55,35){1.8}
\Vertex(45,23){1.8}
\end{picture}
\begin{picture}(92,90)(0,0)
\SetWidth{1.2}
\SetColor{Black}
\ArrowLine(5,35)(25,35)
\ArrowLine(25,35)(45,35)
\ArrowLine(45,35)(65,35)
\ArrowLine(65,35)(85,35)
\SetColor{Black}
\COval(45,62)(12,12)(0){Yellow}{Yellow}
\Text(45,64)[t]{\tiny \rm QCD}
\SetWidth{1.2}
\SetColor{Black}
\Photon(45,74)(45,84){1.8}{2.5}
\SetColor{Green}
\Vertex(45,75){1.8}
%
\SetColor{Red}
\Photon(45,50)(45,35){2}{3}
\Vertex(45,50){1.8}
\Vertex(45,35){1.8}
%
\SetColor{Black}
\SetColor{Black}
\COval(45,15)(12,12)(0){Yellow}{Yellow}
\Text(45,17)[t]{\tiny \rm QCD}
%
\SetColor{Black}
\Photon(25,35)(35,23){2}{3}
\SetColor{Green}
\Vertex(25,35){1.8}
\Vertex(35,23){1.8}
%
\SetColor{Red}
\Photon(65,35)(55,23){2}{3}
\SetColor{Red}
\Vertex(65,35){1.8}
\Vertex(55,23){1.8}
\end{picture}
\\
\begin{picture}(92,90)(0,0)
\SetWidth{1.2}
\SetColor{Black}
\ArrowLine(5,35)(20,35)
\Line(20,35)(25,35)
\ArrowLine(25,35)(85,35)
\SetColor{Black}
\COval(20,62)(12,12)(0){Yellow}{Yellow}
\Text(20,64)[t]{\tiny \rm QCD}
\SetWidth{1.2}
\SetColor{Black}
\Photon(20,74)(20,84){1.8}{2.5}
\SetColor{Green}
\Vertex(20,75){1.8}
%
\SetColor{Red}
\Photon(20,50)(20,35){2}{3}
\Vertex(20,50){1.8}
\Vertex(20,35){1.8}
%
\SetColor{Black}
\SetColor{Black}
\COval(55,15)(12,12)(0){Yellow}{Yellow}
\Text(55,17)[t]{\tiny \rm QCD}
%
\SetColor{Black}
\Photon(35,35)(45,23){2}{3}
\SetColor{Green}
\Vertex(35,35){1.8}
\Vertex(45,23){1.8}
%
\SetColor{Red}
\Photon(75,35)(65,23){2}{3}
\SetColor{Red}
\Vertex(75,35){1.8}
\Vertex(65,23){1.8}
\end{picture}
\begin{picture}(92,90)(0,0)
\SetWidth{1.2}
\SetColor{Black}
\ArrowLine(5,35)(20,35)
\Line(20,35)(25,35)
\ArrowLine(25,35)(85,35)
\SetColor{Black}
\COval(20,62)(12,12)(0){Yellow}{Yellow}
\Text(20,64)[t]{\tiny \rm QCD}
\SetWidth{1.2}
\SetColor{Black}
\Photon(20,74)(20,84){1.8}{2.5}
\SetColor{Green}
\Vertex(20,75){1.8}
%
\SetColor{Red}
\Photon(20,50)(20,35){2}{3}
\Vertex(20,50){1.8}
\Vertex(20,35){1.8}
%
\SetColor{Black}
\SetColor{Black}
\COval(55,15)(12,12)(0){Yellow}{Yellow}
\Text(55,17)[t]{\tiny \rm QCD}
%
\SetColor{Red}
\Photon(35,35)(45,23){2}{3}
\SetColor{Red}
\Vertex(35,35){1.8}
\Vertex(45,23){1.8}
%
\SetColor{Black}
\Photon(75,35)(65,23){2}{3}
\SetColor{Green}
\Vertex(75,35){1.8}
\Vertex(65,23){1.8}
\end{picture}
\begin{picture}(92,90)(0,0)
\SetWidth{1.2}
\SetColor{Black}
\ArrowLine(5,35)(25,35)
\ArrowLine(25,35)(45,35)
\ArrowLine(45,35)(65,35)
\ArrowLine(65,35)(85,35)
\SetColor{Black}
\COval(45,62)(12,12)(0){Yellow}{Yellow}
\Text(45,64)[t]{\tiny \rm QCD}
\SetWidth{1.2}
\SetColor{Black}
\Photon(45,74)(45,84){1.8}{2.5}
\SetColor{Green}
\Vertex(45,75){1.8}
%
\SetColor{Red}
\Photon(45,50)(45,35){2}{3}
\Vertex(45,50){1.8}
\Vertex(45,35){1.8}
%
\SetColor{Black}
\SetColor{Black}
\COval(45,15)(12,12)(0){Yellow}{Yellow}
\Text(45,17)[t]{\tiny \rm QCD}
%
\SetColor{Red}
\Photon(25,35)(35,23){2}{3}
\SetColor{Red}
\Vertex(25,35){1.8}
\Vertex(35,23){1.8}
%
\SetColor{Black}
\Photon(65,35)(55,23){2}{3}
\SetColor{Green}
\Vertex(65,35){1.8}
\Vertex(55,23){1.8}
\end{picture}
\caption{Summary of $O(\alpha^3)$ unwanted diagrams.
}
\label{fig:summary:unwantedDiagram}
\end{center}
\vspace*{-\intextsep} 
\end{figure}
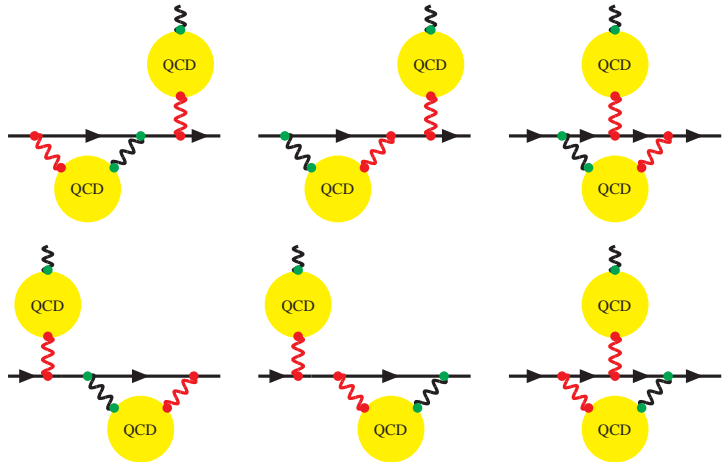

 We discuss how subtraction of unwanted contributions
is realized by $\left(\textcolor[named]{Red}{-\mathcal{K}_{D}}\right)$.
 For that purpose, it may be helpful to observe
the situation by focusing on a $\left(2_E,\,2\right)$-type diagram.
 It is generated in three ways as shown in Fig.~\ref{fig:(2E,2)}\,;
each diagram in Fig.~\ref{fig:(2E,2)}
contains the unwanted contribution, i.e. $O(\alpha^3)$ HVP contribution.
 Note that the HVP contribution coming from 
$\textcolor[named]{Blue}{\mathcal{M}_C}$ can be canceled 
by that supplied from $\textcolor[named]{Blue}{\mathcal{S}_C}$, 
but the other two survive.
 The essential difference between them is as follows.
 The HVP contribution canceled 
by $\textcolor[named]{Blue}{\mathcal{S}_C}$, 
$\textcolor[named]{Green}{\mathcal{S}_{C^\prime}}$
or $\textcolor[named]{Magenta}{\mathcal{S}_{D}}$ 
contains one HVP function entirely supplied from QED average
(QCD average of fully red quark loop),
but the uncanceled one does not.
 The same is true for the other disconnected-type diagrams.
 If the full HVP function is denoted as in Fig.~\ref{fig:fullHVP},
the unwanted contributions that survive 
in the absence of the last term in Eq.~(\ref{eq:nonperturbativeQEDMethod})
can be summarized in Fig.~\ref{fig:summary:unwantedDiagram}, 
where each diagram of identical topology turns out
to {\it appear exactly twice}.
 One can show that 
{\it 
a set of the $O(\alpha^3)$-diagrams generated
by $\textcolor[named]{Red}{\mathcal{K}_{D}}$ 
exactly coincides with that in Fig.~\ref{fig:summary:unwantedDiagram} 
with the same degeneracy}, 
verifying that subtraction is realized in the nonperturbative QED method.

\section{Summary}

 Here, we remarked that, 
to avoid double counting in the prediction of the muon $g-2$,
we must explicitly subtract 
$O(\alpha^3)$ HVP contributions
in every method
for the computation of full HLbL contribution.
 We presented an idea (\ref{eq:nonperturbativeQEDMethod})
for the concrete method, 
which is based on the dynamical (${\rm QCD} + {\rm QED}$) simulation 
as done in Ref.~\cite{Borsanyi:2014jba},
though the results in Ref.~\cite{Jin:2015eua} 
encourage to develop alternative methods
without stochastic realization of virtual photons.

\acknowledgments
T.B is supported by U.S.~DOE grant \#DE-FG02-92ER41989.
N.H.C and L.C.J are supported by U.S.~DOE grant \#de-sc0011941.
M.H is supported by Grants-in-Aid for Scientific Research \#25610053.
T.I and C.L are supported by U.S.~DOE Contract \#AC-02-98CH10996(BNL).
T.I is also supported by Grants-in-Aid for Scientific Research \#26400261.


\end{document}